\documentclass[fleqn,12pt,twoside]{article}
\usepackage{espcrc1}

\usepackage{graphicx}
\usepackage{epsfig}
\usepackage{psfig}

\def\Journal#1#2#3#4{{#1} {\bf #2}, #3 (#4)}

\def\NPA{{\em Nucl.~Phys.} A}
\def\PLB{{\em Phys.~Lett.}  B}

\def\PRC{{\em Phys.~Rev.} C}
\def\ZPC{{\em Z.~Phys.} A}
\def\ZPC{{\em Z.~Phys.} C}

\def\PRep{\em Phys.~Rep.~}

\newcommand{\expl}{\langle \!\langle}
\newcommand{\expr}{\rangle \!\rangle}
\newcommand{\AmS}{{\protect\the\textfont2
  A\kern-.1667em\lower.5ex\hbox{M}\kern-.125emS}}

\title{A novel scenario for the production of
antihyperons
\\
in relativistic heavy ion collisions}

\author{Carsten Greiner\address{Institut f\"ur Theoretische Physik I,
Universit\"at Giessen, D-35392 Giessen, Germany}
\thanks{This work was supported by
the Bundesministerium f\"ur Bildung und Forschung and by the
Gesellschaft f\"ur Schwerionenforschung.}}
       
\begin{document}

\maketitle

\begin{abstract}
We elaborate on our recent suggestion on antihyperon
production in relativistic heavy ion collisions
by means of multi-mesonic (fusion-type) reactions.
It will be shown that
the (rare) antihyperons are driven
towards chemical equilibrium with pions, nucleons and kaons
on a timescale of 1--3 fm/c in a still moderately
baryon-dense hadronic environment.
\vspace*{5mm}
\end{abstract}

\begin{figure}[htp]
\centerline{\psfig{figure=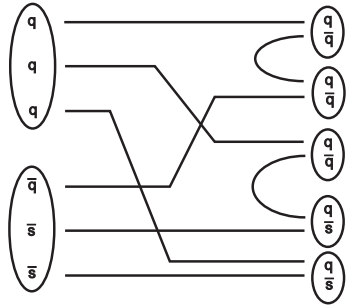,width=4.5cm}}
\caption{Schematic picture for the reaction
$\bar{\Xi } + N \leftrightarrow 3 \pi  + 2  K$.} \label{fig:hyperon}
\end{figure}

The original idea behind the collective enhancement of strangeness
out of a potentially deconfined state of matter, the quark gluon plasma
(QGP),
is that the strange (and antistrange) quarks are
thought to be produced more easily and hence also more abundantly
as compared to the production via
highly threshold suppressed inelastic hadronic collisions.
In addition, the analysis of measured
abundancies of hadronic particles
within thermal models supports the idea of
having established a (thermodynamically) equilibrated hadronic fireball
in some late stage of the reaction
(for analyses of Pb+Pb collisions at CERN-SPS see \cite{BHS99,Redlich}).
In this respect especially a nearly fully chemically equilibrated yield
of strange antibaryons, the antihyperons, had originally
been advocated as an appropriate QGP signature \cite{KMR86}.
Although intriguing, this may not be the correct (or only) interpretation
of the observed antihyperon yields:
In the following we will
elaborate on our recent idea \cite{GL00}
of rapid antihyperon production by multi-mesonic reactions like
$n_1\pi + n_2 K \rightarrow \bar{Y}+p $
(see fig.~\ref{fig:hyperon} for a particular illustration).
As we will demonstrate, this might indeed explain
the observed excess of antihyperons.

Before, let us first remind that
nonequilibrium inelastic hadronic reactions
can explain to a very good extent the overall strangeness production
seen experimentally \cite{Ge98}.
The major amount
of produced strange particles (kaons, antikaons and $\Lambda $'s)
at SPS-energies
can be understood in terms of early and energetic primary,
secondary and ternary non-equilibrium interactions. Strange quarks are
thought to be produced initially via string excitations.
However, applying the usual concept of binary collisions
within the transport approaches, it had been the original claim \cite{KMR86}
that chemical equilibrium values for the abundancies of
antihyperons can not be explained
by successive binary (strangeness exchange) reactions.
Only by incorporating more exotic descriptions like color-rope formation
\cite{So95} or high-dense droplet formation \cite{WA93}
in a hadronic transport model a more abundant production of antihyperons
can be obtained.

If the hadronic degrees of freedom are in a state of thermodynamical
equilibrium, which constitutes the basic concept of thermal
model analyses, a dynamically realization to describe
such a system has to fulfill the concept of detailed balance
in the considered `chemical' reactions.
Hence, multi-mesonic (fusion-like) `back-reactions'
involving $n_1$ pions and $n_2$ kaons of the type
\begin{equation}
\label{antihyp1}
{n}_1 \pi+ n_2  K \,
\leftrightarrow \,
\bar{Y} + N \, \, ,
\, \, \,
\end{equation}
corresponding to the inverse of the strong binary
baryon-antihyperon annihilation process (similar to the standard
baryon annihilation $\bar{p} + p \rightarrow n \, \pi $),
have, in principle, to be taken care of in a dynamical simulation.
In (\ref{antihyp1}) $n_2$ counts
the number of anti-strange quarks within the antihyperon $\bar{Y}$.
${n}_1+ n_2$ is expected to be around $5-7 $.

\begin{figure}[htb]
\begin{minipage}[t]{78mm}
\psfig{figure=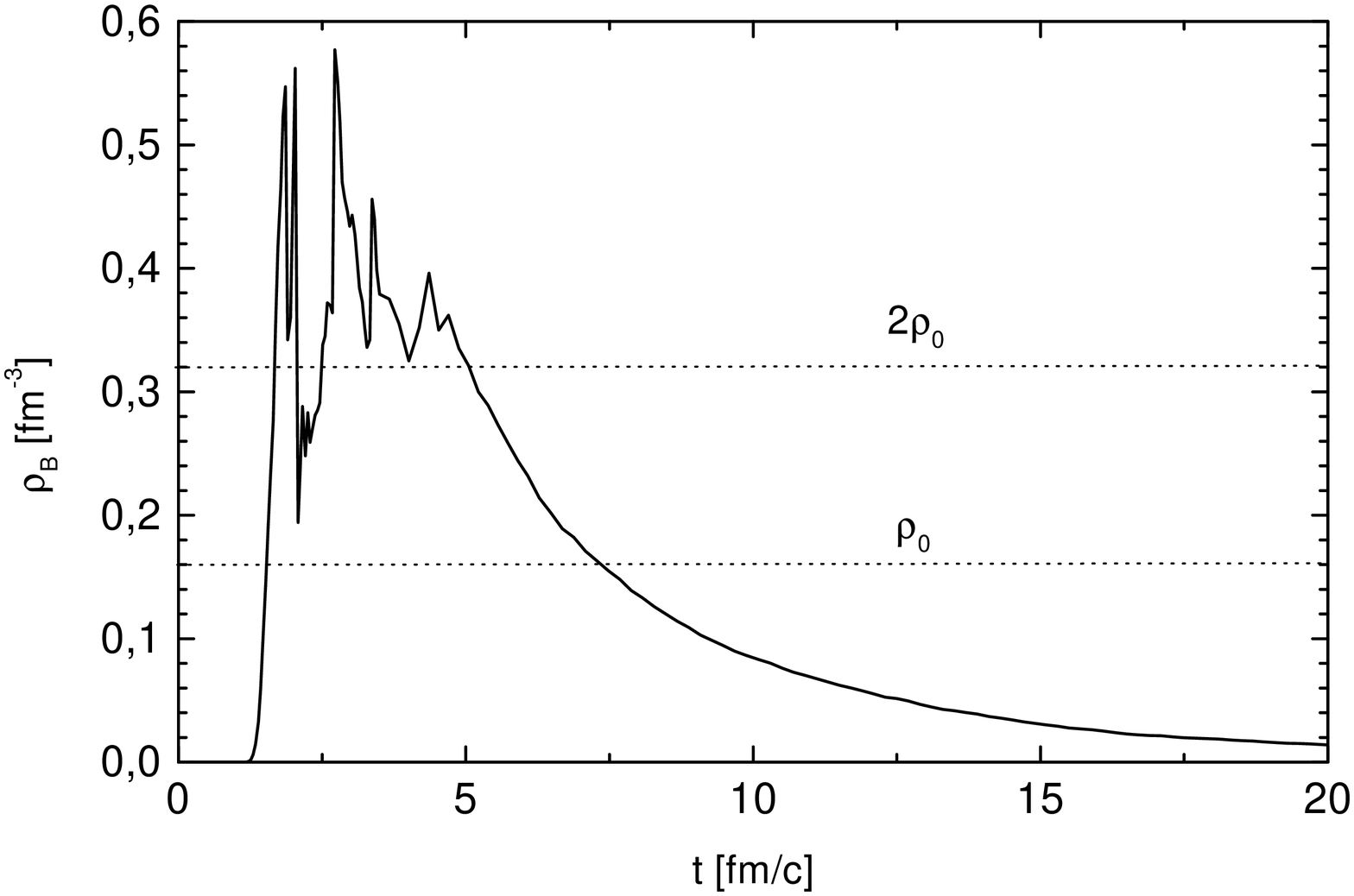,width=87mm}
\caption{Time evolution of the (average) net baryon density
at midrapidity $|\Delta Y| \leq 1$ and central Pb+Pb-collision.
The amount of baryon number residing still in string-like
excitations is explicitely discarded.}
\label{fig:WC}
\end{minipage}
\hspace{\fill}
\begin{minipage}[t]{78mm}
\psfig{figure=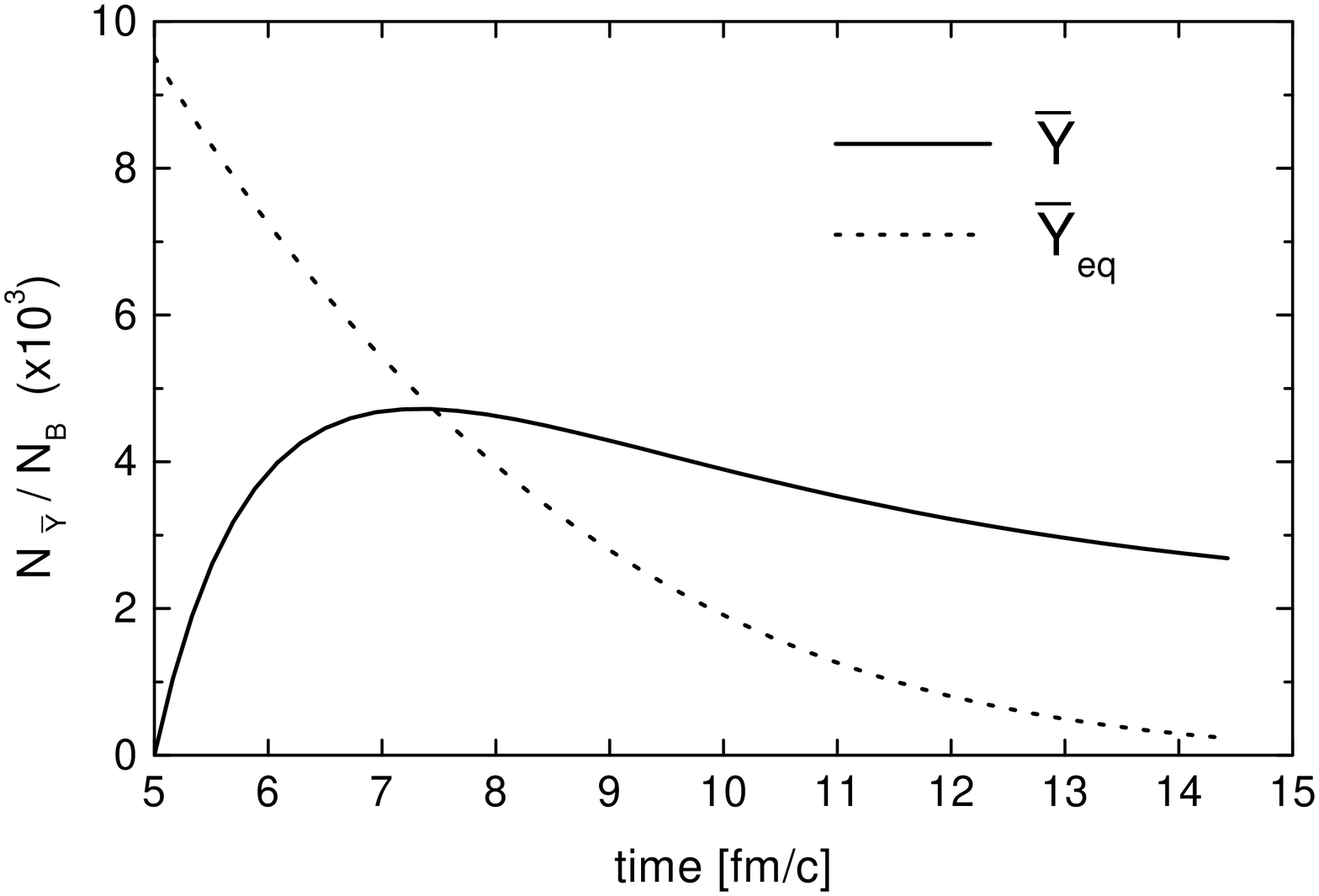,width=87mm}
\caption{The solution for $N_{\bar{Y}}/N_B \,  (t)$ and
$N_{\bar{Y}}^{eq.}/N_B \,  (t)$ for an isentropic expansion.
$N_{\bar{Y}}(t_0=5\, \mbox{fm/c})$ is set to zero.
}
\label{fig:vlin}
\end{minipage}
\end{figure}

Furthermore, the annihilation rate $\Gamma _{\bar{Y}}$ is expected to be large:
It is plausible to assume that the
annihilation cross sections are approximately similar in magnitude as
for $p+\bar{p}$
at the same relative momenta. Hence, with
$\sigma _{p \bar{Y}\rightarrow n_1 \pi + n_2 K}
\approx 50 $ mb one has
$(\Gamma  _{\bar{Y}})^{-1} \equiv 1/
(\expl \sigma _{\bar{Y}N} v _{\bar{Y}N} \expr \rho_N) \approx 1-3 \, fm/c $,
when adopting for the baryon density $ \rho_N  \approx 1-2 \rho_0 $.
The latter value seems reasonable: In fig.~\ref{fig:WC}
we have depicted the (net) baryon density as a function of time
at a space region for particles at midrapidity obtained within the
microscopic transport model HSD as employed in \cite{Ge98}.
The figure illustrates
that a pure hadronic fireball (without any
string-like excitations) at a (net) baryon density $\rho_B \approx 2\rho_0$
has started to be established after approximately 5 fm/c after the onset of the
collision.

The annihilation together with the multi-mesonic reactions
do effectively lead to the following master equation for
the yield of a specific antihyperon specie
\begin{equation}
\label{mastera}
\frac{d}{dt} N _{\bar{Y}} \, =\,   -
\Gamma _{\bar{Y}} \left\{
N _{\bar{Y}} \,
-  \,
N _{\bar{Y}} ^{eq.}
\sum_{{n_1}}\, p_{n_1}
\left( \frac{N_\pi}{N_\pi^{eq.}} \right) ^{n_1}
\left( \frac{N_K}{N_K^{eq.}} \right) ^{n_2}
\right\} \, \, \, .
\end{equation}
Here $p_{n_1}$ states the (unknown) relative probability
of the reaction (\ref{antihyp1}) to decay into a specific number $n_1$ of pions
with $\sum_{n_1} p_{n_1} =1$. This master equation can be derived from
an underlying Boltzmann-type transport equation \cite{CG01}.
The second term describes
the gain per unit time in the hyperon yield due to several
coalescing pions and kaons.
(The microscopic picture of this complicated process is
completely hidden; the term is dictated by the principle of detailed
balance.)
Assuming further that
the pions, baryons and kaons do stay in thermal and chemical
equilibrium, the master equation becomes simply
\begin{equation}
\label{masterd}
\frac{d}{dt} N _{\bar{Y}} \,  = \,    - \,
\Gamma  _{\bar{Y}}
\left\{
N _{\bar{Y}} \, -  \,
N ^{eq }_{\bar{Y}}
\right\} \, \, \, .
\end{equation}
The annihilation rate thus dictates the timescale
for (local) chemical equilibration of the antihyperons
together with the pions, nucleons and kaons.
This scale competes
with the expansion timescale of the late hadronic fireball, which
is in the same range or larger. In any case
these multi-mesonic, hadronic reactions
can explain a sufficiently
fast equilibration before or at the (so called) chemical freeze-out
(with
parameters given by the thermal model analyses,
where the (net) baryon density has dropped to around 0.5 - 1 $\rho_0 $
\cite{BHS99,Redlich}).
This interpretation does rest on the (conservative) view
that before the chemical freeze-out already a
hadronic system has established.

To be more quantitative we show in fig.~\ref{fig:vlin}
the number of one particular specie of
antihyperons (normalized to the net baryon number)
as a function of time when solving the above master equation.
The initial abundancy is set to zero.
The pions, kaons and nucleons are assumed to stay in
thermodynamical equilibrium throughout a (late-stage) isentropic expansion
with parameters characteristic for SPS energies.
The net baryon density
$\rho_B(t)$
as a function of time has been effectively parametrized
from fig.~\ref{fig:WC}. Further details and scenarios
will be given in \cite{CG01}.
In addition, the instantaneous equilibrium abundancy of antihyperons,
$N_{\bar{Y}}^{eq.}(T(t),\mu_B(t),\mu_s(t))/N_B$, is also depicted. This
equilibrium number strongly decreases as a function of time (or decreasing
temperature).
At a time of 8-10 fm/c (3-5 fm/c after the onset of the description)
the antihyperons effectively do `freeze' out
roughly at their particular equilibrium value
(at a temperature around $150-160$ MeV) at that
given time interval.  Refering to fig.~\ref{fig:WC}
this happens at a baryon density of
around 0.5 - 1 $\rho_0 $ and which coincides with the parameters of the
thermal model analyses.
Beyond that `point' at
moderately low baryon densities
the multi-mesonic
creation processes
becomes more and more ineffective
because of the rapid expansion.
This might then also
explain the `position' of the chemical freeze-out point
for the antihyperons \cite{CG01}.
In this respect it would also be very interesting to
adopt the reasoning in the recent work by Rapp and Shuryak \cite{RS00}
(which, in part, had actually triggered our work), that the pions and also
now the kaons might in fact go out of perfect chemical equilibrium below
chemical freeze-out, as their total abundancies effectively
have to stay constant.

To summarize,
multi-mesonic production of antihyperons is a consequence
of detailed balance and, as the rate $\Gamma_{\bar{Y}}$ is
indeed very large, seems to be the by far most dominant source
in a hadronic gas.
This is a remarkable observation,
as it clearly demonstrates
the importance of hadronic multi-particle channels.
At the moment such `back-reactions' are not included
in the present transport codes and some new strategy
has to be invented \cite{CG01a}
in order to be more competitive for a direct
comparison with various experimental findings.
Also, our argument should perfectly apply at AGS energies.
Indeed a large ratio of
$\bar{\Lambda }/\bar{p} \approx 2-3 $ for some central
rapidity window has been observed by various collaborations, which
might indeed also favor a scenario of nearly
chemically saturated strange and nonstrange antibaryon
populations.
More precise experimental data
as well as a theoretical analysis along the line of thermal models
would be most welcome.
Measurements of antihyperon production
could also be done at possible future heavy-ion facilities
at GSI.
This would
be a very interesting opportunity to unreveal the mechanism here proposed.
\\[5mm]
{\bf Acknowledgments}
\\[3mm]
This work has been done in part with S.~Leupold, with whom the idea
of multi-mesonic antihyperon production has been developed.
The author is also indebted to W.~Cassing for providing fig.~\ref{fig:WC}
and for detailed and valuable comments.
He also wants to thank U.~Mosel
for his continuing interest in this particular subject on the
(possible) importance of multi-particle interactions in
heavy-ion collisions.


\begin{thebibliography}{99}
\bibitem{BHS99} P.~Braun-Munzinger, I.~Heppe and J.~Stachel,
\Journal{\PLB}{465}{1}{1999};
\\
F.~Becattini et al, hep-ph/0002267.
\bibitem{Redlich} K.~Redlich, contribution to these proceedings.
\bibitem{KMR86} P.~Koch, B.~M\"uller and J.~Rafelski,
\Journal{\PRep}{142}{167}{1986}.
\bibitem{GL00} C.~Greiner and S.~Leupold, nucl-th/0009036; C.~Greiner,
nucl-th/0011026.
\bibitem{Ge98} J.~Geiss, W.~Cassing and C.~Greiner,
\Journal{\NPA}{644}{107}{1998}.
\bibitem{So95} H.~Sorge, \Journal{\ZPC}{67}{479}{1995};
\Journal{\PRC}{52}{3291}{1995}.
\bibitem{WA93} K.~Werner and J.~Aichelin, \Journal{\PLB}{308}{372}{1993}.
\bibitem{CG01} C.~Greiner, in preparation.
\bibitem{RS00} R.~Rapp and E.~Shuryak, hep-ph/0008326;
R.~Rapp, contribution to these proceedings.
\bibitem{CG01a} C.~Greiner and W.~Cassing, work in progress.


\end{thebibliography}
\end{document}